\documentclass[twocolumn,aps,prx,showpacs]{revtex4-1}
\usepackage{makeidx}
\usepackage{bm}
\usepackage{amssymb}
\usepackage{amssymb,amsmath}
\usepackage{graphicx}
\usepackage{graphics}
\usepackage{dsfont}
\usepackage{array}
\usepackage{slashed}
\usepackage{bm}

\newcommand{\nn}{\nonumber}

\usepackage{color}

\begin{document}
\title{New quantum transition  in Weyl semimetals with correlated disorder}
\author{T. Louvet, D. Carpentier, and A. A. Fedorenko}
\affiliation{\mbox{Univ Lyon, Ens de Lyon, Univ Claude Bernard, CNRS, Laboratoire de Physique, F-69342 Lyon, France}}

\date{\today}


\begin{abstract}

A Weyl semimetal denotes an electronic phase of solids in which two bands cross linearly. In this paper we study the effect of a
spatially correlated disorder on such a phase. Using a renormalization group analysis, we show that in three dimensions, three
scenarios are possible
depending on the disorder correlations.
A standard transition is recovered for short range correlations.
For disorder decaying slower than $1/r^{2}$, the Weyl semimetal  is unstable to any weak disorder and no transition persists.
In between, a new phase transition occurs. This
transition still separates a disordered metal from a semi-metal, but with a new critical behavior that we analyze to two-loop order.

\end{abstract}

\maketitle

\section{Introduction}

The Anderson localization transition is the seminal exemple of a disorder driven quantum phase transition between a metallic and an insulating
phase~\cite{Evers:2008}. Recently,  a different transition has attracted a lot of attention, in which disorder drives a zero gap semi-conductor with linear band crossing
 into a diffusive metal (for review see~\cite{Syzranov:2016b}).

 In this second case,
the density of state~(DOS) at the band crossing is actually increased by the disorder, as opposed to the standard situation
 for the Anderson transition.
Interest in these phases which we will call semimetals
has been steadily developing for several years, fueled by intense activity on both theoretical and experimental aspects. It is now well known that relativistic massless fermions emerge as 2D
electronic excitations in graphene and as surface states of 3D topological insulators.
Two years ago, the successive discovery of materials hosting linear band crossings in 3D  has revived the excitement around the field~\cite{Xu:2015a,Xu:2015b}.
Indeed,  the 3D case is special in that a linear crossing between two bands, called a Weyl point, possesses a topological property which encodes its protection against gap opening by perturbations preserving translational invariance.
Quite naturally,  a weak scalar disorder, which is inescapably present in real materials, is expected to preserve the band crossing of these phases.

 Such a stability with respect to disorder can be cast into the renormalization group (RG) framework.
 While in two dimensions any amount of disorder is relevant and destabilizes the band crossing,
a weak disorder is irrelevant for three dimensional fermions with linear dispersion relation.
The crossing band point persists in the presence of weak quenched disorder, which
results only in a non-universal renormalization of the Fermi velocity. Thus, a weakly  disordered  system
behaves qualitatively as a clean  sample:  the density of states (DOS) vanishes quadratically with energy,
up to exponentially small corrections due to rare events~\cite{Nandkishore:2014,Pixley:2016}.
It will also feature pseudoballistic transport properties~\cite{Sbierski:2014}
and a vanishing zero-frequency optical conductivity~\cite{Roy:2016b}.
However, a strong enough  disorder drives the system from a semimetal
to a disordered  metallic phase
through a transition which is different from the Anderson one.
This disordered metal is characterized by  a finite DOS at zero energy~\cite{Fradkin:1986},
but its precise nature, \textit{e.g.}  Anderson insulator or diffusive metal,
depends on the precise nature of the phase, {\it e.g.} Dirac versus Weyl semimetals or number of cones~\cite{Altland:2015,Altland:2016,Syzranov:2015,Syzranov:2015c,Garttner:2015}.
This transition has been recently intensively studied
both numerically~\cite{Sbierski:2015,Goswami:2011,Kobayashi:2014,Chen:2015}
as well as 
analytically using $2+\varepsilon$ ~\cite{Syzranov:2015b,Roy:2014}
and $4-\varepsilon$~\cite{Louvet:2016,Pixley:2016b} expansions.
In particular, the transition is characterized through the critical exponents $\nu$ and $z$:
 at this transition, the correlation length diverges as $\xi~\sim |\Delta-\Delta_c|^{-\nu} $ where $\Delta_c$ is the critical disorder strength,
 while the dynamic critical exponent $z$ is defined trough the scaling  between energy and momenta  at the transition $\omega \sim k^z$.
Besides, it was found that exactly at the transition the wave functions and local DOS exhibit multifractality similar to the Anderson transition but
 with different universality classes~\cite{Louvet:2016,Syzranov:2016}.

It is well known that correlations of the disorder potential, overlooked in this previous studies, can be  present experimentally and change the nature of the
transition. This is indeed the case for the Anderson transition~\cite{Croy:2011}. Moreover, the low energy properties of the Dirac phase in graphene
are known to be sensitive to disorder correlations \cite{Fedorenko:2012} .
Such correlations may originate from the presence of linear dislocations, planar grain boundaries, unscreened charge impurities, etc.

In the present paper we consider the effects of long-range disorder correlations on the semimetal-diffusive metal transition
in the  Weyl semimetals restricting consideration to a single cone.

The paper is organized as follows. Section~\ref{sec:model} introduces the model which is renormalized  in Sec.~\ref{sec:ren}.  We study the phase diagram and calculate the critical exponents to two-loop order in Sec.~\ref{sec:exponents} and conclude in Sec.~\ref{sec:conclusion}.

\section{Model} \label{sec:model}

Relativistic fermions moving in a $d$-dimensional space in the presence  of an external potential $V(r)$ can be described by the euclidian  action
\begin{eqnarray} \label{eq:action0}
&& \!\!\!\!\!\!\!\! S =
\int d^d r \int d \tau \bar{\psi}(\mathbf{r},\tau)(\partial_\tau  - i  \gamma_{j} \partial^{j}   + V(\mathbf{r})  )
\psi(\mathbf{r},\tau),
\end{eqnarray}
where $\psi$ and $\bar{\psi}$ are independent Grassmann fields  and  $ \tau$  is the imaginary time.
The three-dimensional (3D) Weyl fermions corresponds to $\gamma_j = \sigma_j$, $j=1,2,3$ given by the Pauli matrices.
In general the $\gamma_j$ are elements of a Clifford algebra
  satisfying the anticommutation relations:
$\gamma_i \gamma_j + \gamma_j \gamma_i = 2 \delta_{ij} \mathbb{I}$, and $i,j=1,...,d$.
We assume that the disorder potential $V(\mathbf{r})$ is a random
Gaussian variable with zero mean and the variance
$\overline{V(\mathbf{r})V(\mathbf{r}')}=g(\mathbf{r}-\mathbf{r}')$,  $g(r) \sim r^{-a}$. For convenience,
we fix the normalization of the  variance in the Fourier space
\begin{eqnarray} \label{eq:dis-cor}
\tilde{g}(k) = \Delta_1+\Delta_2 k^{a-d},
\end{eqnarray}
 which must be positive.  The $\Delta_1$ term  in Eq.~(\ref{eq:dis-cor}) corresponds to  the short-range (SR)
disorder since  it becomes the Dirac $\delta$-function in real space: the impact of such  uncorrelated  disorder has been studied previously 
in  Refs.~\cite{Sbierski:2015,Goswami:2011,Kobayashi:2014,Syzranov:2015b,Roy:2014,Chen:2015,Louvet:2016,Pixley:2016b}.
The strength of the long-range  (LR) correlated disorder is given by $\Delta_2$.
To average over disorder we use the replica trick and introduce $N$ copies of the
original system. After averaging over disorder we arrive at the replicated effective action
\begin{eqnarray}
\!\!\!\!\!  S_{\mathrm{eff}} &=&
\int d \tau  d^d r  \bar{\psi}_{\alpha}(\mathbf{r},\tau)(\partial_\tau - i  \gamma_{j} \partial^{j}   )
  \psi_{\alpha}(\mathbf{r},\tau) \nonumber \\
 &&  - \frac{1}2
  \int d\tau_1  d\tau_2 d^d r_1 d^d r_2  g(\mathbf{r}_1-\mathbf{r}_2)  \nonumber \\
&&
  \times  \bar{\psi}_{\alpha}(\mathbf{r},\tau_1) \psi_{\alpha}(\mathbf{r},\tau_1)
\bar{\psi}_{\beta}(\mathbf{r},\tau_2) \psi_{\beta}(\mathbf{r}, \tau_2), \ \ \ \ \ \ \label{eq:action1}
\end{eqnarray}
where summation over repeated replica indices $\alpha,\beta=1,...,N$ is implied.
The properties of the original system with quenched disorder can be obtained
by taking the limit $N\to 0$. Introducing the Matsubara frequency $\omega$ the action~(\ref{eq:action1})
can be rewritten in the Fourier space as
\begin{eqnarray} \label{eq:action-b}
&&  S = \int_{k,\omega} \bar{\psi}_{\alpha}(-\mathbf{k},-\omega)(  \bm{\gamma} \mathbf{k}  -
 i \omega  )
  \psi_{\alpha}(\mathbf{k},\omega) \nonumber \\
&& \ \ \ - \frac{ 1}{2} \int_{k_i,\omega_i} \,
\tilde{g}\left(\mathbf{k}_1+\mathbf{k}_2\right)
\bar{\psi}_{\alpha}(\mathbf{k}_1,-\omega_1) \psi_{\alpha}(\mathbf{k}_2,\omega_1) \nonumber \\
&& \ \ \  \times
\bar{\psi}_{\beta}(\mathbf{k}_3,-\omega_2) \psi_{\beta}(-\mathbf{k}_1-\mathbf{k}_2-\mathbf{k}_3,\omega_2), \ \ \ \
\end{eqnarray}
where $\int_k := \int \frac{d^d k}{(2\pi)^d} $ and $\int_\omega := \int \frac{d \omega }{(2\pi)} $.
We now define the  correlation functions with insertions of the composite operator $\mathcal{O} (r) := \bar{\psi}_{\alpha}(r){\psi}_{\alpha}(r)$,

\begin{align} \label{eq:correlation-funs}
& G^{(2n,l)}(\{x\},\{y\},\{z\})
 \nn \\
& = \left\langle \bar{\psi}_{\alpha_1}(x_1) \psi_{\alpha_1} (y_1) ...\bar{\psi}_{\alpha_{n}}(x_{n}) \psi_{\alpha_{n}} (y_{n})  \mathcal{O}(z_1)...\mathcal{O}(z_l) \right\rangle,
\end{align}
where we used the shortcut notation $x_i:= (r_i, \tau_i)$.
For example, the local DOS can be found from the retarded Green function
\begin{equation}
\rho(E) =  - \frac1{\pi} \mathrm{Im}  G^{R}(r,r, E), \label{eq:DOS-def}
\end{equation}
which is related to  $G^{(0,1)} (r, \omega )$ by the analytic continuation $i\omega \to E + i 0$.
The correlation functions~(\ref{eq:correlation-funs})
can be calculated perturbatively in small $\Delta_1$ and $\Delta_2$.
Each term of this perturbation series can be represented as a Feynman diagram in which  lines
stand for the the bare propagator
\begin{equation}
\langle \bar{\psi}_\alpha(\mathbf{k},\omega) \psi_{\beta}(-\mathbf{k},-\omega) \rangle_0
=\delta_{\alpha\beta}
\frac{\bm{\gamma} \mathbf{k}  + i \omega}{k^2+\omega^2}.
\end{equation}
and there are two types of vertices which correspond to $\Delta_1$ and $\Delta_2$ terms in action~(\ref{eq:action-b}) with
$\tilde{g}(k)$ given by Eq.~(\ref{eq:dis-cor}).
Both of them transmit only momenta but not frequency and the second vertex explicitly depends on the transmitted momenta as $k^{a-d}$.

\section{Renormalization and scaling behavior} \label{sec:ren}

Dimensional analysis shows that weak disorder is irrelevant for $d>2$,
nevertheless the system can undergo a phase transition to a diffusive metal for
strong enough disorder. The correlation
functions~(\ref{eq:correlation-funs}) computed perturbatively in small disorder
turn out to be diverging in $d=2$ which is the lower critical dimension of
the transition. To describe the scaling behavior of the system in the vicinity of
this transition we apply the field-theoretic renormalization group to two-loop order.
To that end we calculate the correlation functions using
dimensional regularization. Following Ref.~\cite{Dudka:2016} we perform
a double expansion  in $d=2+\varepsilon$ and $a=2+\delta$ in such a way that
the UV divergences are converted into the poles in $\varepsilon$  and $\delta$.
In the framework of the minimal subtraction scheme we consider ratios like  $\varepsilon/\delta$ to be finite in the limit
$\varepsilon,\delta \to 0$. We do not include them into the counterterms, choosing the latter to be the pole part only.
The poles in $\varepsilon$ and $\delta$ can be accumulated
in the renormalization factors: $Z_{\psi}$, $Z_\omega$, $Z_{1}$ and $Z_{2}$, so that
the renormalized action can be written as
\begin{eqnarray} \label{eq:action-r}
&&  S_R = \int_{k,\omega} \bar{\psi}_{\alpha}( Z_{\psi}  \bm{\gamma} \mathbf{k}  -
 i Z_{\omega}\omega  )   \psi_{\alpha}\nonumber \\
&& - \frac{ 1}{2} \int_{k_i,\omega_i} \,
\left( \mu^{-\varepsilon } Z_1 \Delta_1 +  \mu^{-\delta } Z_2 \Delta_2 |\mathbf{k}|^{a-d}\right) \bar{\psi}_{\alpha} \psi_{\alpha}\bar{\psi}_{\beta} \psi_{\beta}. \nn \\
\end{eqnarray}
Here we have introduced the renormalized fermionic fields $\psi$, $\bar{\psi}$
and the renormalized dimensionless  coupling constants $\Delta_1$ and $\Delta_2$
on the mass scale $\mu$. In what follows we will denote the bare variables by a ring. The renormalized variables
are related to the bare ones by
 \begin{eqnarray} \label{eq:Z-factors}
&&\mathring{\psi} = Z_{\psi}^{1/2} \psi, \ \ \ \mathring{\bar{\psi}} = Z_{\psi}^{1/2}  \bar{\psi},\\
&&\mathring{\omega_j} = Z_\omega Z_{\psi}^{-1}\omega_j, \ \ \ \  \mathring{O} = Z_\omega Z_\psi^{-1} O, \\
&& \mathring{\Delta}_1 =
 \frac{2 \mu^{-\varepsilon }}{K_d} \frac{Z_1}{Z_{\psi}^{2}} \Delta_1, \ \ \ \
  \mathring{\Delta}_2 =
 \frac{2 \mu^{-\delta }}{K_d} \frac{Z_2}{Z_{\psi}^{2}} \Delta_2. \label{eq:delta-ren}
 \end{eqnarray}
For the sake of simplicity
we have also included $K_d/2$ into definition of the renormalized coupling constants in Eqs.~(\ref{eq:delta-ren}).
$K_d = 2\pi^{d/2}/((2\pi)^d\Gamma(d/2))$ is the surface area of the $d$-dimensional unite sphere divided by $(2\pi)^d$
coming from the angular integration in the Feynmann diagrams.
The renormalized and the bare Green functions are related by
\begin{equation}
  \mathring{G}^{(2n,l)}(\{p, \mathring{\omega}\} ,\mathring{\Delta}) =
Z_\omega^{l} Z_{\psi}^{n-l}G^{(2n,l)}(\{p, \omega\}, \Delta, \mu), \label{eq-gam2}
\end{equation}
where $\Delta:=\{\Delta_1, \Delta_2\}$.
Using that the bare Green functions   $\mathring{G}^{(2n,l)}$  do not depend on the renormalization scale
$\mu$ one can derive the RG flow equation
\begin{eqnarray}
 &&\left[ \sum\limits_j p_j
\frac{\partial}{\partial p_j} + (1+ \gamma(\Delta)) \sum\limits_j\omega_j \frac{\partial}{\partial \omega_j} \right. \nn \\
&&  \ \ \ + \sum\limits_{i=1,2} \beta_i(\Delta)
\frac{\partial}{\partial \Delta_i}  +d( 2n-1) - n(d-1+\eta_\psi(\Delta))  \nn \\
&& \ \ \
 + l   (1+\gamma (\Delta) ) \Big] G^{( 2n,l)}(\{p, \omega\}, \Delta, \mu)=0, \label{eq-rg1-4-2}
\end{eqnarray}
where we have defined the scaling functions
\begin{eqnarray}
&&\beta_i(\Delta)= - \left.\mu\frac{\partial \Delta_i}{\partial \mu} \right|_{\mathring{\Delta}}, \label{eq:beta}  \\
&&\eta_\psi(\Delta)= -\sum\limits_{i=1,2}\beta_i(\Delta)\frac{\partial \ln Z_\psi}{\partial \Delta_i}, \label{eq:eta} \\
&&\eta_\omega(\Delta)= -\sum\limits_{i=1,2} \beta_i(\Delta)\frac{\partial \ln Z_\omega}{\partial \Delta_i}, \\
&& \gamma(\Delta) = \eta_\omega(\Delta)- \eta_\psi(\Delta).
\end{eqnarray}
The solutions of Eq.~(\ref{eq-rg1-4-2}) can be found by using the method of characteristics.
The characteristics are lines in the space of $p_i$, $\Delta$, $\omega_i$ along which Eq.~(\ref{eq-rg1-4-2}) can be rewritten as
 an ordinary differential equation of the first order.
The characteristics lines can be  parameterized by the auxiliary parameter $\xi$ which will be later identified with the correlation length.
The lines are given by equations:
\begin{eqnarray}
&&   \frac{d p_j (\xi)}{d \ln \xi} = p_j(\xi),  \\
&&   \frac{d \Delta_i (\xi)}{d \ln \xi} = \beta_i(\Delta(\xi)), \label{eq:RG-flow} \\
&&   \frac{d \omega_j (\xi)}{d \ln \xi} = [1+\gamma(\Delta(\xi))] \omega_j(\xi),
\end{eqnarray}
with the initial conditions $\Delta (1)=\Delta$, $p_j(1)=p_j$, and $\omega_j (1)=\omega_j$.
The solution of Eq.~(\ref{eq-rg1-4-2}) propagates along the characteristics lines
according to the ordinary differential equation equation
\begin{eqnarray}
\frac{d  \ln H_{ 2n,l}(\xi)}{d \ln \xi} &=& d( 2n-1) -  n ( d-1 + \eta_\psi(\Delta(\xi)) )  \nn \\
&&  + l   (1+\gamma (\Delta)).
\end{eqnarray}
with the initial conditions $H_{ 2n,l}(1)=1$. Thus the solution of Eq.~(\ref{eq-rg1-4-2})
satisfies
\begin{eqnarray}
&& G^{(2n,l)}(p_j, \omega_j, \Delta) = H_{2n,l}(\xi) G^{( 2n,l)}( p_j(\xi),\omega_j(\xi), \Delta(\xi) ).  \nn \\ \label{eq-rg4-2}
\end{eqnarray}
Since the DOS is related to $G^{(0,1)}$ by Eq.~(\ref{eq:DOS-def}) it obeys the scaling relation
\begin{eqnarray}
&& \rho (\omega, \Delta) = H_{0,1}(\xi) \rho ( \omega (\xi) , \Delta(\xi) ). \ \ \ \label{eq-rg4-22}
\end{eqnarray}
Let us now assume that the RG flow (\ref{eq:RG-flow}) has a fixed point (FP) $\Delta^*=(\Delta_1^*,\Delta_2^*)$
defined as
\begin{eqnarray}
\beta_i(\Delta^*)=0, \ \ \  i = 1,2.  \label{eq:fixpoint0}
\end{eqnarray}
To determine the stability properties of the FP one can linearize the flow equation in its vicinity computing the stability matrix
\begin{equation}\label{smatrix}
\mathcal{M}_{ij} = \left. \frac{\partial \beta_i(\Delta)}{\partial \Delta_j}\right|_{\Delta^*},
\end{equation}
which has two eigenvalues $\lambda_1$ and $\lambda_2$.
The  positive eigenvalues of the matrix~(\ref{smatrix}) correspond to the unstable directions in the plane $(\Delta_1,\Delta_2)$.
The transition if it exists is controlled by a FP which has only one of two directions unstable, e.g. $\lambda_1>0$ and $\lambda_2<0$ .
Introducing  the  eigenvector $\Delta-\Delta^*$ associated with $\lambda_1>0$
the scaling formula~(\ref{eq-rg4-2}) in the vicinity of the FP~(\ref{eq:fixpoint0}) can be written as

\begin{eqnarray}
  &&\!\!\!\! G^{(2n,l)}  (p_i, \omega_i,\Delta)   \nn \\
  && \ \ \ = \xi^{d(2n-1) - 2n d_\psi + l z} g_{2n,l}( p_i \xi ,\omega_i \xi^z, |\Delta-\Delta^*| \xi^{1/\nu}). \nn \\
\end{eqnarray}
Here  $|\Delta-\Delta^*|$ is the length of the eigenvector and
we have defined the critical exponents for  the  correlation length $\xi$,
\begin{eqnarray}
\xi \sim |\Delta-\Delta^*|^{-\nu}, \ \ \ \ \frac1{\nu}=\lambda_1,
\end{eqnarray}
the dynamic critical exponent
 \begin{eqnarray}
\omega \sim k^z, \ \ \ \ \ z = 1+\gamma(\Delta^*) \label{eq:def-z}
\end{eqnarray}
and  the anomalous dimension of the fields $\psi$ and $\bar{\psi}$
\begin{eqnarray}
d_\psi = \frac12[d-1+\eta_\psi(\Delta^*)].
\end{eqnarray}
For instance, the two-point correlation function, 
which gives the momentum
distribution at the transition,  behaves as
\begin{eqnarray}
\!\!\! G^{( 2,0)}(p) \sim p^{-1+\eta_\psi(\Delta^*)}, \ \  G^{( 2,0 )}(r) \sim \frac1{r^{d-1+\eta_\psi(\Delta^*)}}.\ \ \ \ 
\end{eqnarray}
The scaling formula~(\ref{eq-rg4-22}) for the DOS  in the vicinity of the FP~(\ref{eq:fixpoint0})  has the form

\begin{eqnarray}
 && \rho( \omega) = \xi^{z-d} \rho_0(\omega \xi^z, |\Delta-\Delta^*| \xi^{1/\nu}). \ \ \
\end{eqnarray}

\section{Transitions: existence and critical behavior} \label{sec:exponents}

\begin{figure}
\includegraphics[width=75mm]{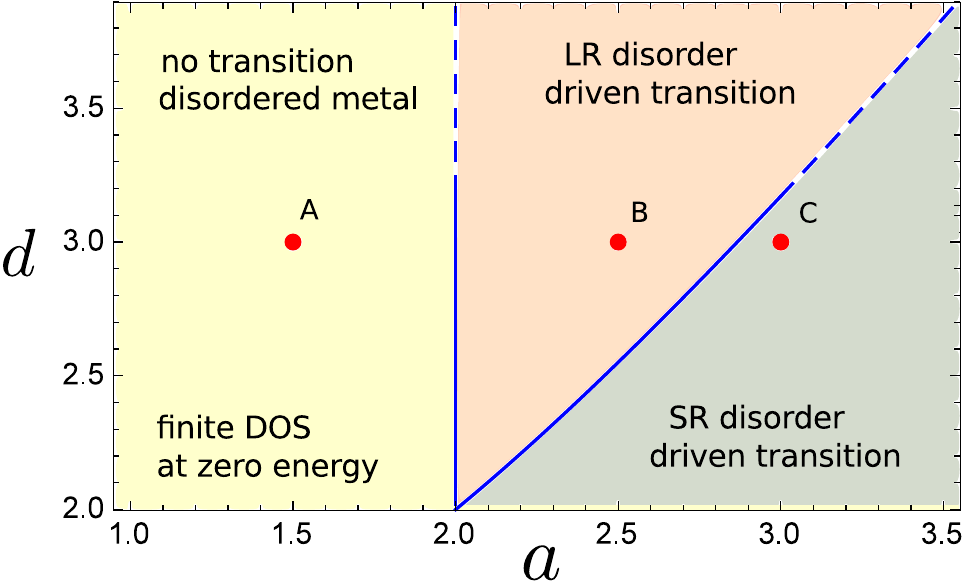}
\caption{Domains of existence of the transition as a function of  $a$ and  $d$: (i) in the left region there is no transition, the disorder is always relevant  and the  disordered metal is the only phase. The typical RG flow  computed at the point A [$a=3/2$, $d=3$]  is
shown in  Fig.~\ref{fig:flow-NO}; in the middle region there is a new transition from a semimetal phase to a metallic phase controlled by the LR FP. The typical RG flow  computed at the point B [$a=5/2$, $d=3$]  is shown in  Fig.~\ref{fig:flow-LR}; in the right region there is a transition from a semimetal phase to a metallic phase controlled by the SR FP. The typical RG flow  computed at the point C [$a=3$, $d=3$] is shown in  Fig.~\ref{fig:flow-SR}.
 }
  \label{fig:phase-diagram}
\end{figure}

\begin{figure}
\includegraphics[width=75mm]{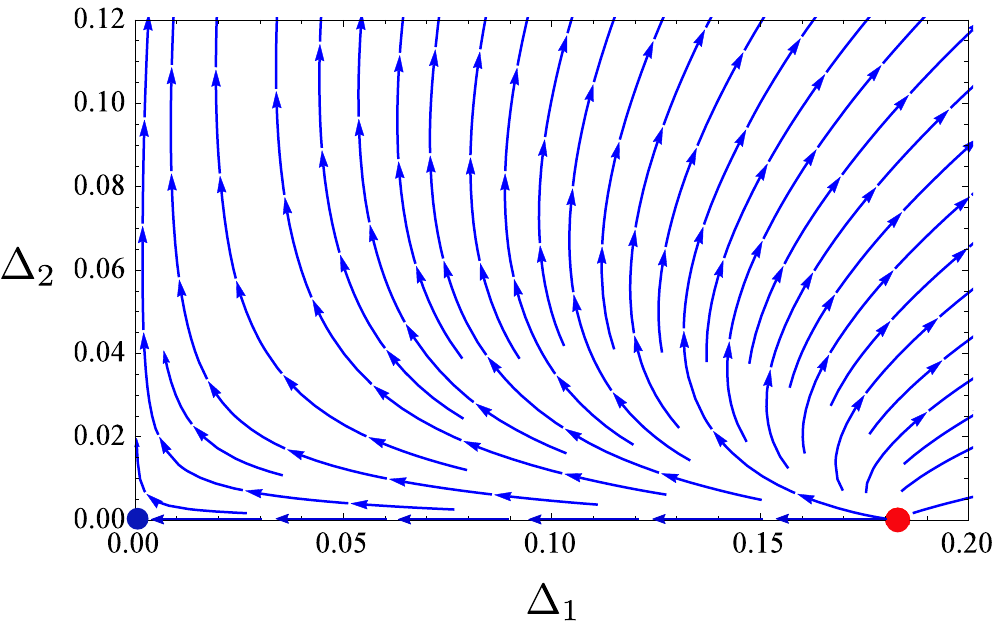}
\caption{ {\it Strongly disordered phase:} runaway of the RG flow for $d=3$ ($\varepsilon=1$) and $a=3/2$ ($\delta=-1/2$) corresponding to the point A in Fig.~~\ref{fig:phase-diagram}.
The red dot is the SR FP which is fully unstable. The blue dot is the Gaussian FP corresponding to the semimetal phase.}
  \label{fig:flow-NO}
\end{figure}

\begin{figure}
\includegraphics[width=75mm]{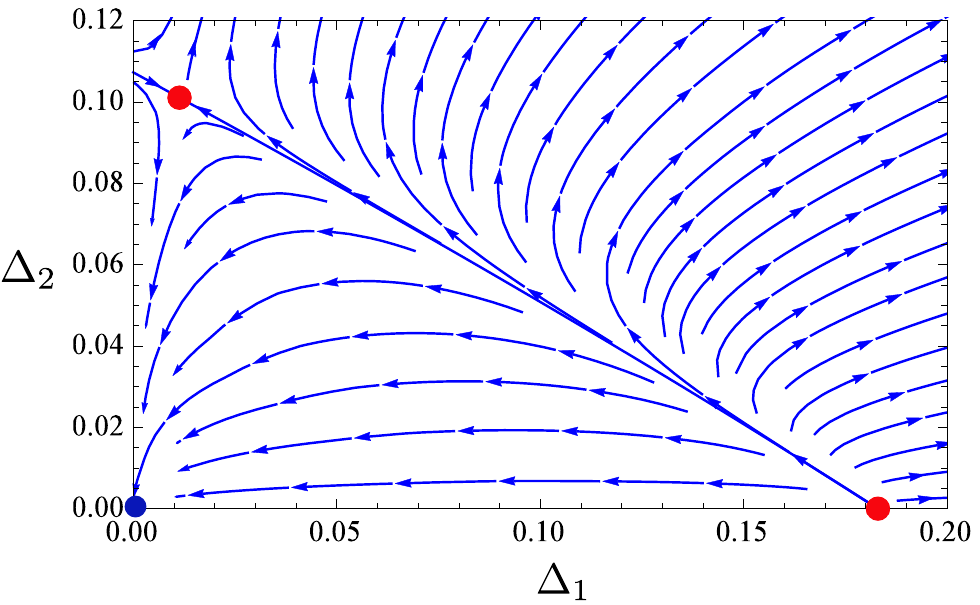}
\caption{{\it New LR transition:} the RG flow for $d=3$ ($\varepsilon=1$) and $a=5/2$ ($\delta=1/2$)  corresponding to the point B in Fig.~~\ref{fig:phase-diagram}.
The right red dot is the SR FP which is fully unstable.   The left red dot is the  new LR FP
which controls the transition between a semimetal and a disordered metal phase
and which gives rise to a new universality class different from the SR case.
The blue dot is the Gaussian FP corresponding to the semimetal phase.  }
  \label{fig:flow-LR}
\end{figure}

Using the two-loop diagrams computed in Ref.~\cite{Dudka:2016} we derive the beta functions \eqref{eq:beta}:
\begin{eqnarray}
 \beta_1(\Delta_1,\Delta_2) &=& -\varepsilon \Delta_1 + 4\Delta_1^2 + 4 \Delta_1\Delta_2 + 8\Delta_1^3  \nn \\
 && +20 \Delta_1^2 \Delta_2  + 4 \Delta_2^3 +16 \Delta_1 \Delta_2^2, \label{eq:RGflow-1} \\
 \beta_2(\Delta_1,\Delta_2) &=& -\delta \Delta_2 + 4\Delta_2^2 + 4 \Delta_1\Delta_2 + 4\Delta_2^3  \nn \\
 && + 4 \Delta_1^2 \Delta_2  + 8\Delta_1 \Delta_2^2, \label{eq:RGflow-2}
\end{eqnarray}
the dynamic critical exponent \eqref{eq:def-z}:
\begin{eqnarray}
 z(\Delta_1,\Delta_2) &=& 1+ 2(\Delta_1 + \Delta_2) + 2(\Delta_1 + \Delta_2)^2, \label{eq:RGflow-z}
\end{eqnarray}
and the anomalous dimension \eqref{eq:eta} of the fermionic fields
\begin{eqnarray}
 \eta_{\psi}(\Delta_1,\Delta_2) = -2 \Delta_1^2+2 \Delta_2^2  -\frac{4  \varepsilon }{\delta }\Delta_2 (\Delta_1 + \Delta_2).  \ \ \label{eq:eta-psi-exp}
\end{eqnarray}
Note that the ratio $\varepsilon/\delta$ in Eq.~(\ref{eq:eta-psi-exp}) 
is finite in our regularization  scheme.

\begin{figure}
\includegraphics[width=75mm]{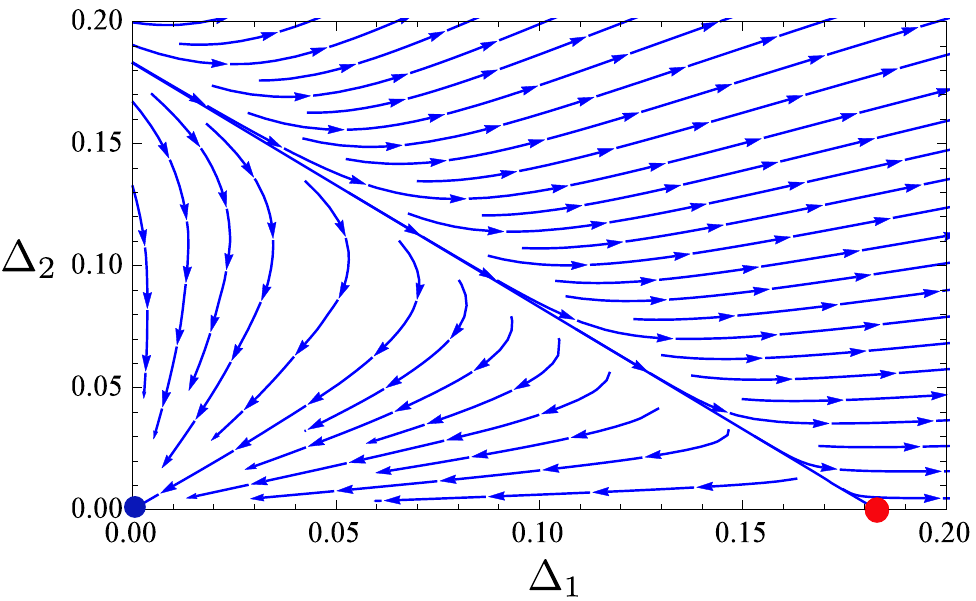}
\caption{ {\it SR transition}: the RG flow for  d=3 ($\varepsilon=1$) and $a=3$ ($\delta=1$)  corresponding to the point C in Fig.~~\ref{fig:phase-diagram}.
The red dot is the SR FP  with a single unstable direction (relevant operator), describing the semimetal to disordered metal transition governed by the parameter $\Delta_1$. The blue dot is the Gaussian FP corresponding to the semimetal phase.}
  \label{fig:flow-SR}
\end{figure}

We now analyze the RG flow derived from Eqs.~(\ref{eq:RGflow-1})-(\ref{eq:eta-psi-exp}) that is summarized in Figures~\ref{fig:phase-diagram}-\ref{fig:flow-SR}.
We are interested in the effect of an additional LR correlated disorder 
on the SR disorder driven transition:
our results on the stability of the SR fixed point with respect to additional $\Delta_2$ distinguish between three different domains, as shown on Fig.~\ref{fig:phase-diagram}.
First note that the case $d=2$ ($\varepsilon = 0$),
 corresponding to graphene, is special since it corresponds to the lower critical dimension where no transition occurs: SR disorder is
 marginally relevant and drives the system to a strong disordered metallic phase, characterized by a finite zero-energy DOS~ \cite{Fedorenko:2012}.
  For dimension $d$  greater than $2$ we must distinguish three regimes of disorder correlations :
 \begin{itemize}
 \item[(A)]For $a<2$ ($\delta<0$) i.e. when the long-range disorder correlations decay
  slower than $1/r^2$,  the semi-metallic phase becomes unstable to any small amount of disorder.
 This manifests itself into  the instability of the SR fixed point to additional LR disorder:
 the transition is suppressed and the system always flows towards the strong disordered metallic phase as shown in Fig.~\ref{fig:flow-NO}.

\item[(C)]  For the opposite case of "short range"  disorder correlation, defined by an exponent $a$ larger than a
critical value of $a$, $a>a_c(d)$ defined below \eqref{eq:deltac-def}, the SR fixed point is stable and the semimetal to metal SR transition remains unaffected. Fig.~\ref{fig:flow-SR} shows that the relevant direction for the RG flow is still along $\Delta_1$ axis.

\item[(B)] In the intermediate domain  $2 < a < a_c(d)$ we find that in presence of LR disorder a transition
 still exists between a semi-metal and a disordered metal, but is different from the SR transition : this corresponds to the
existence of
 a new LR fixed point of the RG, see Fig.~\ref{fig:flow-LR}.
 Correspondingly, the critical properties
of this new LR disorder-driven transition are different from the previous SR one.
\end{itemize}
Thus, we have found that depending on the dimension and the type of algebraic decay, LR correlated disorder can strongly affect the disorder driven transition in semimetals.
Indeed, depending on the values of $\varepsilon$ and $\delta$ the RG flow equations (\ref{eq:RGflow-1}) and (\ref{eq:RGflow-2})
have up to three FPs:

(i) the Gaussian fixed point (Gaussian FP)  is
\begin{eqnarray}
\Delta_1^\mathrm{G}=\Delta_2^\mathrm{G}=0.
\end{eqnarray}
The basin of attraction of this
FP in the plane $(\Delta_1,\Delta_2)$  corresponds to the semimetal phase. For instance, for $\delta<0$ the basin of attraction collapses to
the axe $\Delta_2=0$  so that the semimetal phase is washed out by any weak correlated disorder.

(ii) the short-range fixed point (SR FP)  reads
\begin{eqnarray}
 && \Delta_1^{\mathrm{SR}} = \frac{1}{4} \left(\sqrt{2 \varepsilon +1}-1\right)
  = \frac{\varepsilon }{4}-\frac{\varepsilon ^2}{8}+O\left(\varepsilon ^3\right), \\
 &&  \Delta_2^{\mathrm{SR}}=0,
\end{eqnarray}
which has a single unstable direction for
$\delta>\delta_{\mathrm{c}}(\varepsilon)$, where
\begin{eqnarray}\label{eq:deltac-def}
\delta_{\mathrm{c}}(\varepsilon) = \frac12(\sqrt{1+2\varepsilon} +\varepsilon-1) \approx \varepsilon -\frac{\varepsilon ^2}{4}+O\left(\varepsilon ^3\right),
\end{eqnarray}
and  fully unstable  otherwise. This defines $a_c(d)= 2+\delta_c(d-2)$. In $d=3$ we find $a_c(3) \approx 2.8 $.

(iii) long-range fixed point (LR FP) reads
\begin{eqnarray}
 \Delta_1^{\mathrm{LR}} &=& \frac{(4+\delta)\sqrt{\delta +1} -3 \delta-4  }{2( \varepsilon -\delta) } \approx \frac{\delta ^3}{16 (\varepsilon-\delta  )}+..., \ \ \ \ \\
 \Delta_2^{\mathrm{LR}} &=&  \frac{ (\varepsilon - 2 \delta -4) \sqrt{\delta +1}    +4 -\varepsilon +4\delta }{2 (\varepsilon - \delta )} \ \ \\ && \approx \frac{\delta }{4} - \frac{\delta ^2 \varepsilon }{16 (\varepsilon-\delta  )}+ ... \ \ \  \
\end{eqnarray}
The LR FP is physical, i.e. it corresponds to a positive $\tilde{g}(k)$ and
has a single unstable direction for $0<\delta<\delta_{\mathrm{c}}(\varepsilon)$.

We now discuss the critical properties at the transition.
For $a<2$ ($\delta<0$) there is a runaway of the RG flow so that the correlated disorder  is always relevant
if present. In this case there is no semimetal phase and the DOS at zero energy is finite (see Fig.~\ref{fig:flow-NO}).
For  $2<a<a_{\mathrm{c}}(d)$ ($0<\delta<\delta_{\mathrm{c}}(\varepsilon)$)
there is  a line of phase transitions separating the semimetal and diffusive metal phases with the critical behavior controlled
by the LR FP (see Fig.~\ref{fig:flow-LR}).  The corresponding critical exponents computed to two-loop order are
\begin{eqnarray}
&& \frac1{\nu_{\mathrm{LR}}} = \delta +\frac{\delta ^2 (2 \delta +\varepsilon )}{4 \varepsilon } + O(\varepsilon^3, \delta^3), \label{eq-exp-nu-lr} \\
&& z_{\mathrm{LR}} = 1+ \frac{\delta}{2} + O(\varepsilon^3, \delta^3) ,  \label{eq-exp-z-lr} \\
&& \eta_{\mathrm{LR}} = -\frac{\delta(2\varepsilon-\delta)}{8} + O(\varepsilon^3, \delta^3). \label{eq-exp-eta-lr}
\end{eqnarray}

Note that the two-loop correction to the dynamic critical 
exponent~(\ref{eq-exp-z-lr}) vanishes. The RG function~(\ref{eq:RGflow-z}) 
used to compute this exponent is related to that for the correlation length 
exponent $\nu_{\mathrm{Ising}}(d=2)$ of the 2D Ising model with correlated random 
bond disorder. Scaling arguments  allow one to 
conjecture~\cite{Weinrib:1983}
that at the transition controlled by the LR FP this 
exponent is exactly $\nu_{\mathrm{Ising}}(d)=2/a$ in any $2\le d<4$.
This is in accordance with the recent  two-loop order study of diluted 
2D Ising model using the $2+\delta$ expansion~\cite{Dudka:2016} in which 
the two-loop 
corrections to $\nu_{\mathrm{Ising}}(d=2)$  vanish similarly to~(\ref{eq-exp-z-lr}). 
On this basis we can also conjecture  that the exact value of the dynamic exponent 
for the semimetal-disordered metal transition controlled by
the LR FP is $z_{\mathrm{LR}}=a/2$.

For $a_{\mathrm{c}}(d)<a$ ($\delta_{\mathrm{c}}(\varepsilon)<\delta$)
there is  also a line of phase transitions separating the semimetal and diffusive metal phases with the critical behavior controlled
by the SR FP (see Fig.~\ref{fig:flow-SR}).
The critical exponents at the SR FP to two-loop order are
\begin{eqnarray}
&&\frac1{\nu_{\mathrm{SR}}} =\varepsilon + \frac{\varepsilon^2}{2} + O(\varepsilon^3), \label{eq-exp-nu-sr}\\
&& z_{\mathrm{SR}} = 1+ \frac{\varepsilon}{2} -\frac{\varepsilon^2}{8} + O(\varepsilon^3) , \label{eq-exp-z-sr}  \\
&& \eta_{\mathrm{SR}} = -\frac{\varepsilon^2}{8} + O(\varepsilon^3).  \label{eq-exp-eta-sr}
\end{eqnarray}
 One expects  that on the borderline between the 
regions with SR and LR criticality (see Fig.~\ref{fig:phase-diagram}) the dynamic critical exponent is a continuous
function, i.e we argue that $z_{\mathrm{SR}}(\varepsilon)=z_{\mathrm{LR}}(\delta_c)=a_c/2$. 
Substituting  the one loop result~(\ref{eq-exp-z-sr}) one again arrives 
at Eq.~(\ref{eq:deltac-def}).

Note that our derivation of the critical properties based on a $2+\varepsilon$ 
expansion is valid up to two loop order. Indeed, we have recently 
showed that the consistent description of the transition with SR disorder 
beyond two loops involves an infinite number of relevant operators generated
by the RG flow.  To overcome this obstacle we proposed an alternative way
based on a $4-\varepsilon$ expansion~\cite{Louvet:2016}.
However, generalization of this approach to the case of LR disorder 
is a non trivial task which remains to be done.

\section{Conclusions} \label{sec:conclusion}

We have studied the effect of LR disorder correlations on the semimetal -   disordered metal transition. We have found that
for  slowly decaying correlations $a<2$ the LR correlated disorder is always relevant and drives the system to a diffusive phase:  the transition is suppressed and the DOS is finite at zero energy
for arbitrary weak disorder.
 Let us note that in particular  this result restricts the range of disorder correlations that can be used in simulations
 (necessary to uncouple the different Weyl cones) to actually observe the SR transition.
 More surprisingly, we have found an intermediate regime $2<a<a_c(\varepsilon)$ where the system undergoes a  new transition from semimetal to diffusive metal
at finite disorder strength and the criticality is controlled by  a LR FP. This change of criticality under addition of LR disorder correlations is
a completely new phenomena distinct from what was known for graphene. It would be of great interest to explore the physical consequences of this crossover, e.g. on the behaviour of the DOS at finite energy along the lines of~\cite{Fedorenko:2012}.
Another interesting direction for future studies would be possible instantons occuring for rare disorder realisations, which are expected to decay algebraically~\cite{Nandkishore:2014}.

\acknowledgments We acknowledge  support from the French Agence Nationale de la Recherche through
Grants No. ANR-12-BS04-0007 (SemiTopo).


\bibliographystyle{apsrev4-1}
\bibliography{disorderedWeyl}

\end{document}